\documentclass[sigconf]{acmart}
\AtBeginDocument{%
  }

\copyrightyear{2026}
\acmYear{2026}
\setcopyright{cc}
\setcctype{by}
\acmConference[KDD '26]{Proceedings of the 32nd ACM SIGKDD Conference on Knowledge Discovery and Data Mining V.2}{August 09--13, 2026}{Jeju Island, Republic of Korea}
\acmBooktitle{Proceedings of the 32nd ACM SIGKDD Conference on Knowledge Discovery and Data Mining V.2 (KDD '26), August 09--13, 2026, Jeju Island, Republic of Korea}
\acmDOI{10.1145/3770855.3817796}
\acmISBN{979-8-4007-2259-2/2026/08}
\usepackage[inline]{enumitem}
\usepackage{tikz}
\usetikzlibrary{positioning,fit,arrows.meta,shapes,calc}
\usepackage[ruled,vlined,linesnumbered]{algorithm2e}
\usepackage{multirow}

\begin{document}

\title{UniRank: End-to-End Domain-Specific Reranking of Hybrid Text-Image Candidates}

\author{Yupei Yang}
\authornote{This work was done when the author was a research intern at Alibaba Group.}
\affiliation{%
  \institution{Shanghai Jiao Tong University}
  \city{Shanghai}
  \country{China}
}
\affiliation{%
  \institution{Alibaba Group}
  \city{Hangzhou}
  \country{China}
}
\email{yupei_yang@sjtu.edu.cn}
% \email{zhihe.yyp@alibaba-inc.com}

\author{Lin Yang}
\affiliation{%
  \institution{Alibaba Group}
  \city{Hangzhou}
  \country{China}
}
\email{ziyu.yl@alibaba-inc.com}

\author{Wanxi Deng}
\affiliation{%
  \institution{Alibaba Group}
  \city{Beijing}
  \country{China}
}
\email{wanxi.dengwx@alibaba-inc.com}

\author{Lin Qu}
\affiliation{%
  \institution{Alibaba Group}
  \city{Hangzhou}
  \country{China}
}
\email{xide.ql@taobao.com}

\author{Shikui Tu}
\authornote{Corresponding author.}
\affiliation{%
  \institution{Shanghai Jiao Tong University}
  \city{Shanghai}
  \country{China}
}
\email{tushikui@sjtu.edu.cn}

\author{Lei Xu}
\affiliation{%
  \institution{Shanghai Jiao Tong University}
  \city{Shanghai}
  \country{China}
}
\email{leixu@sjtu.edu.cn}

\renewcommand{\shortauthors}{Yupei Yang et al.}

\begin{abstract}
  Reranking is a critical component in many information retrieval pipelines. Despite remarkable progress in text-only settings, multimodal reranking remains challenging, particularly when the candidate set contains \emph{hybrid} text and image items. A key difficulty is the modality gap: a text reranker is intrinsically closer to text candidates than to image candidates, leading to biased and suboptimal cross-modal ranking. Vision-language models (VLMs) mitigate this gap through strong cross-modal alignment and have recently been adopted to build multimodal rerankers. However, most VLM-based rerankers encode \emph{all} candidates as images, and treating text as images introduces substantial computational overhead. Meanwhile, existing open-source multimodal rerankers are typically trained on general-domain data and often underperform in domain-specific scenarios.
  To address these limitations, we propose \textsc{UniRank}, a VLM-based reranking framework that natively scores and orders \emph{hybrid} text-image candidates without any modality conversion. Building on this hybrid scoring interface, \textsc{UniRank} provides an end-to-end domain adaptation pipeline that includes:
  \begin{enumerate*}[label=(\arabic*)]
    \item an instruction-tuning stage that learns calibrated cross-modal relevance scoring by mapping label-token likelihoods to a unified scalar score; and
    \item a hard-negative-driven preference alignment stage that constructs in-domain pairwise preferences and performs query-level policy optimization through reinforcement learning from human feedback (RLHF).
  \end{enumerate*}
  Extensive experiments on scientific literature retrieval and design patent search demonstrate that \textsc{UniRank} consistently outperforms state-of-the-art baselines, improving Recall@1 by 8.9\% and 7.3\%, respectively.
\end{abstract}

\begin{CCSXML}
<ccs2012>
   <concept>
       <concept_id>10002951.10003317.10003338.10003343</concept_id>
       <concept_desc>Information systems~Learning to rank</concept_desc>
       <concept_significance>500</concept_significance>
       </concept>
 </ccs2012>
\end{CCSXML}

\ccsdesc[500]{Information systems~Learning to rank}

\keywords{Multimodal Reranking; Vision-Language Models; Domain-Specific Fine-Tuning; Hybrid Text-Image Candidates}


\maketitle

\section{Introduction}
Information retrieval (IR)~\cite{baeza1999modern,manning2008introduction,chowdhury2010introduction,hambarde2023information} aims to identify and return the most relevant items from a large corpus in response to a user query. Modern IR systems are commonly built as a multi-stage pipeline, where \emph{embedding-based retrieval} and \emph{reranking} serve as two fundamental components~\cite{zhu2025large,zhou2025large}. In the first stage, a dense embedding model maps queries and items into a shared vector space, enabling efficient approximate nearest neighbor search to retrieve a small set of potentially relevant candidates from millions of documents. In the second stage, a reranker performs more fine-grained relevance modeling over the candidate set and produces the final ranked list, ensuring that the most relevant results are placed at the top. This retrieve-then-rerank paradigm has delivered substantial improvements in text-only IR and has become a standard practice in many real-world applications~\cite{lewis2020retrieval,gao2023retrieval,edge2024local,fan2024survey}.

\begin{figure*}[t]
\centering
\includegraphics[width=\textwidth]{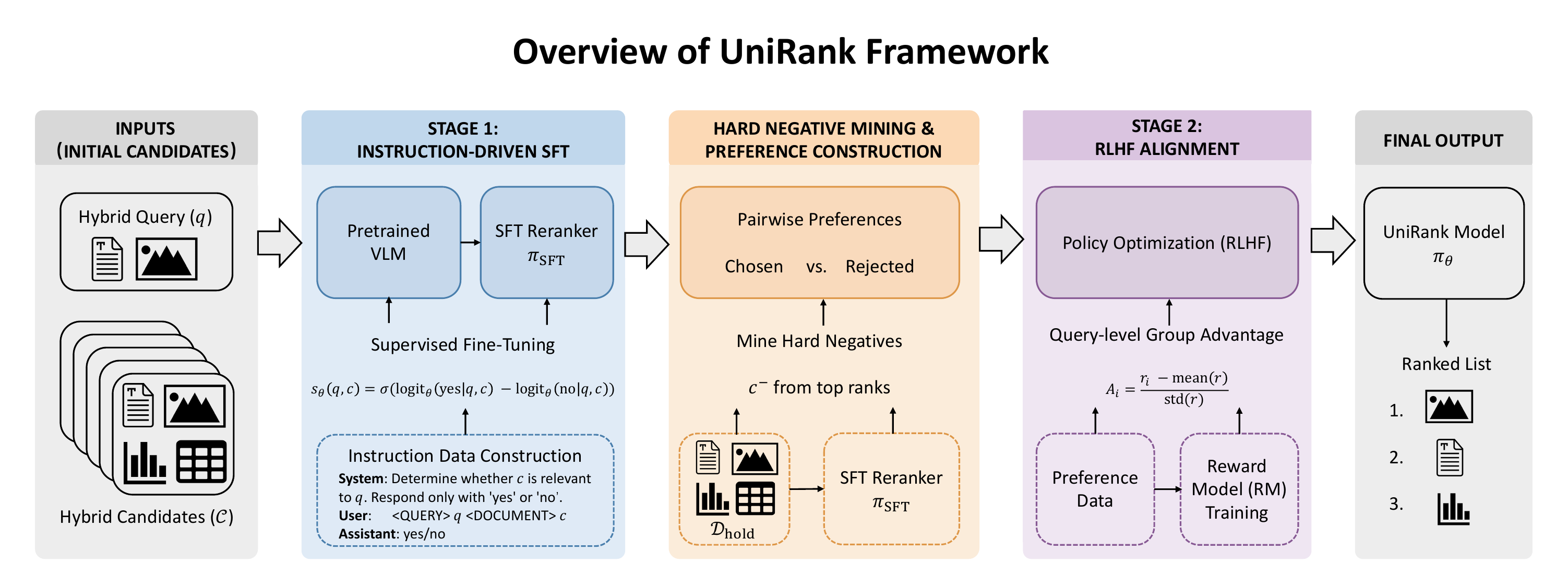}
\caption{Overview of \textsc{UniRank}. We first perform instruction-driven SFT to obtain a VLM-based reranker. We then mine hard negatives to construct pairwise preferences, train a reward model, and further align the reranker via RLHF for domain-specific hybrid text-image reranking.}
\label{fig:unirank-framework}
\end{figure*}

In the past few years, the demand for multimodal retrieval has rapidly grown~\cite{zhao2023retrieving,mei2025survey}, driven by applications such as e-commerce search~\cite{hendriksen2022multimodal}, scientific literature discovery~\cite{wu2024scimmir}, and content recommendation~\cite{yue2025preference}, where both text and images may serve as queries and/or candidates. However, directly extending text rerankers to such settings remains challenging due to the \emph{modality gap}~\cite{geigle2022retrieve,liang2022mind,liu2025benchmarking}: these models are inherently biased toward textual inputs and struggle to fairly compare relevance across heterogeneous modalities. A common workaround is to first convert images into textual descriptions (e.g., captions or VLM-derived text) and then apply a text reranker. While feasible, this approach is fundamentally limited because textual descriptions cannot always faithfully capture the rich and fine-grained semantics of images, leading to a low performance ceiling in practice~\cite{yang2025reid,giouroukis2025s,dong2025mmdocir}. These limitations motivate the need for dedicated multimodal rerankers that can reason over both visual and textual content.

Benefiting from recent advances in VLMs and their strong cross-modal alignment capability, an emerging research direction is to build multimodal rerankers on top of VLMs~\cite{chen2024mllm,xu2025mm,li2026qwen3}. Nevertheless, existing VLM-based rerankers still have notable limitations. First, many approaches homogenize the input format by treating \emph{all} candidates as images so that the reranker can operate in a single visual input space. This design simplifies modeling but incurs substantial storage and inference overhead, and becomes particularly inefficient when the candidate pool is dominated by textual items or requires cross-modal reranking between text and images. Second, while several open-source multimodal rerankers perform well on general-domain benchmarks, they are typically trained on broad web-scale data and lack specialization for domain-specific content. As a result, their performance often degrades when deployed in contexts such as e-commerce, finance, or defense, where private data and unique semantics dominate~\cite{xu2024advancing,jiang2025multimodal,yao2024research}. Crucially, there remains a lack of comprehensive \emph{end-to-end training frameworks} tailored for building domain-adaptive multimodal rerankers—a gap that limits the deployment of high-performance retrieval systems in privacy-sensitive and application-critical scenarios.

To this end, we propose \textsc{UniRank}, an end-to-end framework for training domain-specific rerankers that directly operate on hybrid text-image candidates without modality conversion. Starting from a pretrained VLM backbone, we first construct high-quality instruction-following data and perform supervised fine-tuning (SFT) to obtain a VLM-based reranker that produces meaningful and comparable relevance scores for hybrid candidates. We then run the SFT reranker on held-out in-domain data to mine \emph{hard negatives}---high-scoring but non-relevant candidates---and pair them with gold positives to build large-scale pairwise preference data. Finally, we leverage this preference dataset to train a reward model (RM) and further optimize the reranker via RLHF, improving both score calibration and ranking robustness in domain-specific settings.

Our main contributions are summarized as follows:
\begin{itemize}
    \item \textbf{Hybrid-modality VLM reranking.} We introduce \textsc{UniRank}, a VLM-based multimodal reranking framework that natively operates on \emph{hybrid} candidate sets containing both text and images, enabling unified and fair cross-modal relevance scoring without converting text into images, thus reducing storage and inference overhead.
    \item \textbf{End-to-end domain-specific alignment framework.} We propose an end-to-end, two-stage training framework for domain-specific multimodal rerankers, including
    \begin{enumerate*}[label=(\arabic*)]
        \item instruction-driven SFT to learn calibrated cross-modal relevance scoring, and
        \item hard-negative mining followed by RLHF to construct in-domain pairwise preferences for reward modeling and policy optimization.
    \end{enumerate*}
    \item \textbf{Strong empirical results on two real-world domains.} We evaluate on scientific literature retrieval and design patent search, and show that \textsc{UniRank} consistently outperforms state-of-the-art multimodal rerankers, achieving up to 8.9\% and 7.3\% gains in Recall@1, respectively.
\end{itemize}

\section{Related Work}
\subsection{Rerankers}
\paragraph{Text rerankers}
Text reranking has been extensively studied under the retrieve-then-rerank paradigm, where the reranker refines a candidate set returned by a fast first-stage retriever. A dominant line of work adopts the \emph{cross-encoder} architecture, which jointly encodes the query and a candidate document by concatenation and leverages Transformer-based full token-level interactions to estimate relevance, as in bge-rerankers~\cite{bge_embedding}, jina-rerankers~\cite{wang2025jina}, and mxbai-rerankers~\cite{shakir2024boost}. Another representative direction is \emph{late interaction}, exemplified by ColBERT-style models, which encode queries and documents independently and perform efficient token-level matching at inference time~\cite{khattab2020colbert,wang2023colbert}.

\paragraph{Multimodal rerankers}
Multimodal reranking extends the above problem to settings where queries and/or candidates may contain images in addition to text, and the candidate pool can be heterogeneous (e.g., mixed text and image items)~\cite{wang2012multimodal,wen2024multimodal,mortaheb2025re}. Existing approaches can be broadly categorized into two groups. The first group reduces multimodality to text by converting images into textual representations, such as captions and VLM-extracted text, and then applying a text reranker~\cite{yang2025reid,dong2025mmdocir}. While simple and compatible with mature text rerankers, this strategy may lose visual details and can be bottlenecked by the quality of the intermediate descriptions~\cite{giouroukis2025s}. The second group builds rerankers directly on top of VLMs, leveraging their cross-modal alignment to compare relevance between queries and multimodal candidates~\cite{chen2024mllm,faysse2024colpali}. Recent examples include VLM-based rerankers enhanced with chain-of-thought prompting and task-specific fine-tuning for page-level reranking~\cite{xu2025mm}, as well as methods that finetune Qwen3-VL~\cite{Qwen3-VL} to obtain a reranker by framing reranking as a binary classification problem~\cite{li2026qwen3}. Different from these prior efforts, \textsc{UniRank} targets \emph{cross-modal} reranking over \emph{hybrid} text-image candidate sets while emphasizing an end-to-end training framework for \emph{domain-specific} adaptation.

\subsection{Instruction Tuning and Preference Alignment}
\paragraph{Supervised fine-tuning}
Supervised fine-tuning adapts a pretrained language or vision--language model to downstream tasks by training on curated instruction--response pairs~\cite{wei2021finetuned,ouyang2022training,wang2023self,zhou2023instruction}. In the post-instruction-tuning era, SFT has become a standard recipe to elicit desired behaviors such as following task specifications, producing structured outputs, and learning task-specific scoring or classification heads. To reduce training cost and memory footprint, parameter-efficient fine-tuning (PEFT) methods are widely used~\cite{ding2023parameter,han2024parameter}, such as LoRA~\cite{hu2022lora}, QLora~\cite{dettmers2023qlora}, and AdaLoRA~\cite{zhang2023adalora}. These techniques enable fast customization of large models while keeping most backbone parameters frozen, making them particularly appealing for domain adaptation with limited compute or privacy constraints.

\paragraph{Preference alignment and RLHF}
Beyond SFT, reinforcement learning from human feedback is commonly employed to align model outputs with human or application-specific preferences~\cite{bai2022training,sun2024aligning}. Traditional RLHF trains a reward model from pairwise comparisons and then optimizes the policy using reinforcement learning algorithms such as PPO~\cite{schulman2017proximal}. More recent alternatives like DPO~\cite{rafailov2023direct} simplify the pipeline by directly optimizing a policy with preference data, which avoids explicit reward modeling and RL rollouts. In addition, group-based objectives~\cite{shao2024deepseekmath} have been explored to improve stability and sample efficiency when multiple candidates are available. These approaches provide practical tools for improving calibration, robustness, and controllability, and are increasingly adopted for domain-specific alignment where labeled relevance signals are scarce but preference signals can be collected at scale.

\subsection{Multimodal Fusion}
Early progress in visual representation learning was driven by Vision Transformers (ViT)~\cite{dosovitskiy2020image}, which demonstrated that transformer architectures can serve as strong generic backbones for images. Building on such encoders, contrastive vision--language pretraining methods~\cite{radford2021learning} learned aligned image--text embedding spaces at scale, enabling zero-shot recognition and cross-modal retrieval. Subsequent work further improved vision--language alignment and generation capability via image--text pretraining objectives and multimodal denoising/generation~\cite{li2022blip,li2023blip}, making it possible to move beyond embedding alignment toward unified multimodal understanding and generation.

Recent VLMs typically follow a unified architecture that connects a vision encoder to a large language model (LLM) through a modality adapter (e.g., a projection layer or a lightweight Q-Former~\cite{li2023blip}), enabling the LLM to consume visual tokens and perform multimodal reasoning. Representative families include instruction-tuned VLMs such as MiniGPT-4~\cite{zhu2023minigpt}, LLaVA~\cite{liu2023visual}, and the Qwen-VL series~\cite{Qwen-VL,Qwen2-VL,Qwen2.5-VL,Qwen3-VL}, which combine large-scale pretraining with instruction tuning to support general-purpose multimodal dialogue, grounding, and reasoning. These advances provide a strong foundation for multimodal IR components, including rerankers, by offering improved cross-modal alignment and flexible interfaces for scoring and decision making over mixed visual and textual evidence.

\section{Methodology}
In this section, we provide a detailed description of \textsc{UniRank}. Specifically, we first formalize the task of reranking over hybrid text-image candidates and clarify the input/output interface of a VLM-based reranker. We then introduce our two-stage domain-specific training pipeline:
\begin{enumerate*}[label=(\arabic*)]
    \item supervised fine-tuning with instruction-following data to learn calibrated cross-modal relevance scoring; and
    \item preference-based alignment to further improve ranking quality via hard-negative mining, reward model training, and RLHF.
\end{enumerate*}
An overview of the proposed framework is shown in Figure~\ref{fig:unirank-framework}.

\subsection{Problem Formulation}
We study the reranking problem in a hybrid text-image retrieval setting. Given a query $q$ and an initial candidate set $\mathcal{C}=\{c_1,\dots,c_n\}$ returned by a first-stage retriever, the goal of reranking is to produce a permutation of $\mathcal{C}$ such that the most relevant candidates appear at the top. In our setting, each query and candidate is a multimodal item that may contain text, images, or both, requiring unified scoring across heterogeneous inputs.
% We denote the modality of an instance $x$ as $m(x)\in\{\text{text},\text{image}\}$.

A reranker is parameterized by $\theta$ and computes a relevance score for each query--candidate pair:
\begin{equation}
s_{\theta}(q,c_i) \in \mathbb{R}, \quad c_i \in \mathcal{C}.
\end{equation}
The final ranked list is obtained by sorting candidates in descending order of their scores:
\begin{equation}
\pi = \mathrm{sort}\big(\mathcal{C};\, s_{\theta}(q,\cdot)\big),
\end{equation}
where $\pi$ denotes the resulting permutation. When ground-truth relevance labels are available, we use $y_i \in \{0,1\}$ to indicate whether $c_i$ is relevant to $q$. The training objective is to learn $\theta$ such that relevant candidates receive higher scores than non-relevant ones, i.e.,
\begin{equation}
s_{\theta}(q,c^{+}) > s_{\theta}(q,c^{-}), \quad c^{+}\in \mathcal{C}^{+},\; c^{-}\in \mathcal{C}^{-},
\end{equation}
where $\mathcal{C}^{+}$ and $\mathcal{C}^{-}$ are the relevant and non-relevant subsets of $\mathcal{C}$ for query $q$.

In this work, we instantiate $s_{\theta}$ using a VLM that natively supports multimodal inputs. Specifically, we represent the query and candidate in their original modalities and feed them into the VLM with an instruction prompt. The VLM then generates a short judgment (e.g., \texttt{yes}/\texttt{no}) together with token-level likelihoods, from which we derive a scalar relevance score $s_{\theta}(q,c)$. This interface enables a single reranker to score text--text, image--image, and cross-modal pairs with a unified scoring function.

\subsection{Supervised Fine-Tuning}
\subsubsection{Instruction Data Construction}\label{sec:sft-data}
To convert a general-purpose VLM into a reranker, we cast reranking as an \emph{instruction-following judgment} task: given a query $q$ and a candidate $c$, the model is prompted to output a discrete relevance label indicating whether $c$ is relevant to $q$. The label space $\mathcal{Y}$ can be domain-specific in practice (e.g., \texttt{yes}/\texttt{no} for scientific retrieval and \texttt{high-risk}/\texttt{low-risk} for patent search). Without loss of generality, we denote the two possible outputs as $\mathcal{Y}=\{\texttt{yes},\texttt{no}\}$ throughout this paper. During inference, we derive a scalar relevance score from the model's label token logits by applying a sigmoid to the logit difference:
\begin{equation}\label{eq:score}
s_{\theta}(q,c)=\sigma\big(\mathrm{logit}_{\theta}(\texttt{yes}\mid q,c)-\mathrm{logit}_{\theta}(\texttt{no}\mid q,c)\big),
\end{equation}
where $\sigma(\cdot)$ denotes the sigmoid function. This score is then used for standard top-$k$ reranking.
% Accordingly, SFT should consistently elicit a single label token from the model under multimodal inputs, so that the logits are well-defined and comparable across candidates.
% To this end, we adopt a chat-style instruction format that naturally supports interleaved text and images and enforces a constrained output space. Each training instance is represented as a sequence of messages:
% \begin{equation}
% \mathbf{m} = \big[m^{\text{sys}},\, m^{\text{usr}},\, m^{\text{asst}}\big],
% \end{equation}

To this end, we serialize each labeled query--candidate pair $(q,c,y)$ into a chat-style instruction example, which
\begin{enumerate*}
    \item supports interleaved text and images and 
    \item constrains the model to output a single label in $\mathcal{Y}$.
\end{enumerate*}
Specifically, each example is represented as a three-message conversation:
\begin{equation}\label{eq:chat_message}
\mathbf{m} = \big[m^{\text{sys}},\, m^{\text{usr}},\, m^{\text{asst}}\big],
\end{equation}
where
\begin{enumerate*}[label=(\arabic*)]
    \item $m^{\text{sys}}$ specifies the reranking role and restricts the output space to $\mathcal{Y}$;
    \item $m^{\text{usr}}$ provides the query $q$ and candidate content $c$ in their native modalities; and
    \item $m^{\text{asst}}$ is the ground-truth label $y$.
\end{enumerate*}
Figures~\ref{fig:app_science_sft_pref} and \ref{fig:app_patent_sft_pref} in Appendix \ref{app:prompt_examples} illustrate the overall prompt structure used in \textsc{UniRank}, including the concrete form of the system instruction and the constrained label output.

To be specific, the system message instructs the model to act as a \emph{multimodal relevance judge} and to respond with \emph{only one label} from $\mathcal{Y}$, without any explanation. The user message is constructed by concatenating the query and candidate with explicit delimiters:
\begin{equation}
m^{\text{usr}} = \texttt{<QUERY>: } q \;\; \Vert \;\; \texttt{<DOCUMENT>: } c,
\end{equation}
where $\Vert$ denotes string concatenation. If $q$ or $c$ contains images, we place a special placeholder token \texttt{<image>} at the corresponding position in the user message and pass the raw image(s) through the VLM's vision input channel. This design enables a single prompt template to handle text candidates, image candidates, and mixed query--candidate pairs in a unified manner, without converting one modality into the other. In practice, we construct instruction-following data from in-domain relevance annotations. For each query $q$, annotated relevant items form positive pairs $(q,c^{+}, y^{+})$, while non-relevant items are used to form negative pairs $(q,c^{-}, y^{-})$. We then serialize each labeled pair $(q,c,y)$ into the chat format in Eq.~(\ref{eq:chat_message}), producing the final SFT dataset
$\mathcal{D}_{\text{SFT}}=\{ \mathbf{m}_j \}_{j=1}^{N}$.

\subsubsection{Supervised Fine-Tuning Objective and Hold-Out Data Split}
Given the instruction-following dataset $\mathcal{D}_{\text{SFT}}$, we train the VLM reranker by minimizing the standard language modeling loss over the assistant response tokens. Specifically, for each message sequence $\mathbf{m}$, we compute the negative log-likelihood of the ground-truth label $y \in \mathcal{Y}$ conditioned on the system and user messages:
\begin{equation}
    \mathcal{L}_{\text{SFT}}(\theta) = -\log P_{\theta}\big( y \mid m^{\text{sys}}, m^{\text{usr}} \big),
\end{equation}
where $P_{\theta}(y \mid \cdot)$ denotes the model's conditional probability of generating the label token(s) in $m^{\text{asst}}$. In practice, this is implemented as a cross-entropy loss applied only to the output positions corresponding to the label tokens (e.g., \texttt{yes} or \texttt{no}), while masking all other tokens in the sequence. This objective encourages the model to learn calibrated token-level logits that capture domain-specific relevance semantics, which directly underpin the scoring function \(s_{\theta}(q,c)\) defined in Eq.~\eqref{eq:score}.

To support the subsequent preference alignment stage, we reserve a subset of $\mathcal{D}_{\text{SFT}}$ as hold-out data, denoted $\mathcal{D}_{\text{hold}}$. This subset is excluded from SFT optimization and is only used later for constructing pairwise preference data, ensuring that the preference annotations are derived from in-domain examples unseen during SFT, thereby reducing overfitting and improving generalization in the alignment phase.

\subsection{Preference-based Alignment}
\subsubsection{Preference Data Construction via Hard-Negative Mining}
After SFT, we further improve the reranker via preference-based alignment. This requires a pairwise preference dataset that captures \emph{in-domain} failure modes of the SFT model and provides richer supervision than pointwise labels. To this end, we construct preference data on the held-out split $\mathcal{D}_{\text{hold}}$ by mining \emph{hard} query--candidate prompts on which the SFT reranker is most likely to make mistakes.

Recall that each query--candidate pair is serialized into a prompt context
$\mathbf{x}=(m^{\text{sys}}, m^{\text{usr}})$ (Sec.~\ref{sec:sft-data}), and the model is required to output a label $y \in \mathcal{Y}=\{\texttt{yes},\texttt{no}\}$.
For each query $q \in \mathcal{D}_{\text{hold}}$, we score all candidates $c \in \mathcal{C}_q$ using the SFT model and rank them by $s_{\theta}(q,c)$. We then focus on the top-$N$ candidates and select \emph{hard} prompts where the model tends to over-estimate relevance, i.e., non-relevant candidates appearing in the top-$N$:
\begin{equation}
\mathcal{H}_q = \big\{c \in \mathrm{top}\text{-}N(\mathcal{C}_q)\; \big|\; y(q,c)=0 \big\},
\end{equation}
where $y(q,c)\in\{0,1\}$ denotes the ground-truth relevance.

For each hard negative candidate $c^{-}\in\mathcal{H}_q$, we create a pairwise preference instance at the \emph{response} level by keeping the same prompt $\mathbf{x}=(m^{\text{sys}},m^{\text{usr}})$ and contrasting the \emph{correct} label $y^{\star}$ against the \emph{incorrect} one $\bar{y}$:
\begin{equation}
\big(\mathbf{x},\, y^{\star},\, \bar{y}\big), \quad
y^{\star} = \texttt{no},\;\; \bar{y}=\texttt{yes}.
\end{equation}
That is, for hard negatives the preferred response is the ground-truth label (\texttt{no}) and the rejected response is the SFT model's typical error mode (\texttt{yes}). Symmetrically, for positive prompts $(q,c^{+})$ with $y(q,c^{+})=1$, we construct
\begin{equation}
\big(\mathbf{x},\, y^{\star},\, \bar{y}\big), \quad
y^{\star} = \texttt{yes},\;\; \bar{y}=\texttt{no}.
\end{equation}
Aggregating over queries yields the final preference dataset $\mathcal{D}_{\text{pref}}$. This construction directly targets label-level confusions on difficult in-domain prompts, and is well matched to our reranker formulation where relevance is inferred from label token likelihoods.

\subsubsection{Reward Model Training}
Given the preference dataset $\mathcal{D}_{\text{pref}}$, we train a reward model to predict which \emph{response} is preferred under the same prompt. Concretely, the RM shares the same multimodal prompting interface as the reranker. Let $\mathbf{x}=(m^{\text{sys}}, m^{\text{usr}})$ denote the prompt context constructed from a query--candidate pair, and let $y\in\mathcal{Y}$ be a label token sequence. The RM takes $(\mathbf{x}, y)$ as input and outputs a scalar reward
\begin{equation}
r_{\phi}(\mathbf{x}, y)\in \mathbb{R},
\end{equation}
where $\phi$ denotes the RM parameters. Intuitively, $r_{\phi}(\mathbf{x}, y)$ measures how appropriate it is to respond with label $y$ under prompt $\mathbf{x}$, and thus reflects in-domain relevance preferences.

We optimize the RM using a standard pairwise preference loss. For each preference instance $(\mathbf{x}, y^{\star}, \bar{y})\in\mathcal{D}_{\text{pref}}$, where $y^{\star}$ is the preferred (chosen) label and $\bar{y}$ is the rejected label, we minimize:
\begin{equation}
\mathcal{L}_{\text{RM}}(\phi)
= - \mathbb{E}_{(\mathbf{x}, y^{\star}, \bar{y})\sim \mathcal{D}_{\text{pref}}}
\Big[\log \sigma\big(r_{\phi}(\mathbf{x}, y^{\star}) - r_{\phi}(\mathbf{x}, \bar{y})\big)\Big],
\end{equation}
where $\sigma(\cdot)$ denotes the sigmoid function. This loss corresponds to a Bradley--Terry style model~\cite{bradley1952rank} of pairwise preferences and is widely used in RLHF for learning reward functions from comparison data. Minimizing $\mathcal{L}_{\text{RM}}$ encourages the RM to assign higher reward to the preferred label than to the rejected one under the same prompt.

In practice, we initialize the RM from the SFT reranker to preserve domain knowledge and multimodal alignment, and then fine-tune it on $\mathcal{D}_{\text{pref}}$ using the above objective.

\subsubsection{RLHF Training}
With the learned reward model, we further align the reranker via RLHF to directly optimize ranking behavior under in-domain preferences. We initialize the policy from the SFT reranker, i.e., $\pi_{\theta} \leftarrow \pi_{\theta_{\text{SFT}}}$, and treat the reranker as a policy that, given a prompt context $\mathbf{x}=(m^{\text{sys}},m^{\text{usr}})$, generates a label $y\in\mathcal{Y}$. The reward model $r_{\phi}(\mathbf{x},y)$ provides a learned training signal to encourage better calibrated and more robust relevance scoring.

\textsc{UniRank} is compatible with a range of preference-optimization objectives and RLHF algorithms. In our experiments, we adopt \emph{Group Relative Policy Optimization} (GRPO)~\cite{shao2024deepseekmath}. A key design choice is that we form \emph{query-level groups} rather than the standard prompt-level groups used in typical GRPO settings. That is, instead of sampling multiple responses for a single prompt, we exploit the inherent listwise structure of reranking and compare multiple \emph{candidate prompts} that share the same query.

We construct groups directly from the preference data $\mathcal{D}_{\text{pref}}$, which already contains prompt contexts derived from held-out in-domain queries. Let $\mathcal{X}_q$ denote the set of prompt contexts in $\mathcal{D}_{\text{pref}}$ associated with query $q$ (each corresponding to a distinct candidate under $q$). For each query-level group $\{\mathbf{x}_1,\ldots,\mathbf{x}_G\}\subseteq \mathcal{X}_q$, the current policy generates labels $y_i \sim \pi_{\theta}(\cdot\mid \mathbf{x}_i)$ and receives rewards
\begin{equation}
r_i = r_{\phi}(\mathbf{x}_i, y_i), \quad i=1,\ldots,G.
\end{equation}
We normalize rewards within the group to obtain relative advantages:
\begin{equation}
A_i = \frac{r_i - \mu_r}{\sigma_r + \delta}, \quad
\mu_r = \frac{1}{G}\sum_{j=1}^{G} r_j,\quad
\sigma_r = \sqrt{\frac{1}{G}\sum_{j=1}^{G}(r_j-\mu_r)^2},
\end{equation}
where $\delta$ is a small constant for numerical stability. The policy is then updated using a clipped surrogate objective:
\begin{equation}
\mathcal{L}_{\text{GRPO}}(\theta)
= - \mathbb{E}\Bigg[
\frac{1}{G}\sum_{i=1}^{G}
\min\Big(
\rho_i(\theta)\,A_i,\;
\mathrm{clip}(\rho_i(\theta), 1-\epsilon, 1+\epsilon)\,A_i
\Big)
\Bigg],
\end{equation}
where $\rho_i(\theta)=\frac{\pi_{\theta}(y_i\mid \mathbf{x}_i)}{\pi_{\theta_{\text{ref}}}(y_i\mid \mathbf{x}_i)}$ is the policy ratio, $\epsilon$ is the clipping hyperparameter, and $\pi_{\theta_{\text{ref}}}$ is a fixed reference policy (set to $\pi_{\theta_{\text{SFT}}}$) to regularize updates. By contrasting candidates \emph{within the same query}, \textsc{UniRank} directly optimizes the relative ordering behavior that reranking requires, thereby improving robustness on hard in-domain cases.

After RLHF optimization, we use the aligned policy to compute the final relevance score $s_{\theta}(q,c)$ for reranking.

% \begin{algorithm}[t]
% \caption{\textsc{UniRank}}
% \label{alg:unirank}
% \KwIn{In-domain labeled pairs $\mathcal{D}_{\text{SFT}}$; hold-out split $\mathcal{D}_{\text{hold}}$; top-$K$; group size cap $G$.}
% \KwOut{Aligned reranker policy $\pi_{\theta}$.}

% \textbf{Stage I (SFT):} Train $\pi_{\theta_{\text{SFT}}}$ on $\mathcal{D}_{\text{SFT}}\setminus \mathcal{D}_{\text{hold}}$ using $\mathcal{L}_{\text{SFT}}$.

% \textbf{Preference mining:} Run $\pi_{\theta_{\text{SFT}}}$ on $\mathcal{D}_{\text{hold}}$; for each query $q$, collect hard candidates in $\mathrm{Top}\text{-}K$ and build response-level preferences $\mathcal{D}_{\text{pref}}$.

% \textbf{Stage II (RM):} Initialize RM from $\pi_{\theta_{\text{SFT}}}$ and train $r_{\phi}$ on $\mathcal{D}_{\text{pref}}$ using $\mathcal{L}_{\text{RM}}$.

% \textbf{Stage II (RLHF):} Initialize $\pi_{\theta}\leftarrow \pi_{\theta_{\text{SFT}}}$, set reference $\pi_{\theta_{\text{ref}}}\leftarrow \pi_{\theta_{\text{SFT}}}$; form query-level groups from $\mathcal{D}_{\text{pref}}$ (cap each group to size $G$) and optimize $\pi_{\theta}$ with $\mathcal{L}_{\text{GRPO}}$.

% \Return{$\pi_{\theta}$}
% \end{algorithm}

\section{Experiments}
In this section, we evaluate \textsc{UniRank} on two domain-specific multimodal retrieval tasks, including scientific literature retrieval and design patent search, where the reranking candidate set contains \emph{hybrid} text-image items.\footnote{Code is available at \url{https://github.com/CMACH508/UniRank}} Our evaluation is designed to answer the following research questions:

\begin{itemize}
    \item \textbf{RQ1 (Effectiveness in domain-specific hybrid reranking).}
    Does \textsc{UniRank} achieve better reranking performance than strong text-only and VLM-based multimodal baselines in domain-specific settings with hybrid text-image candidates?

    \item \textbf{RQ2 (Robustness across retrieval setups and model scales).}
    How robust is \textsc{UniRank} under different first-stage retrievers (which change the candidate distribution) and different VLM backbone sizes (which change capacity)?  

    \item \textbf{RQ3 (Contribution of each training component).}
    What is the contribution of each component in \textsc{UniRank}'s end-to-end training pipeline to the final reranking quality?
\end{itemize}

\subsection{Setup}
\paragraph{Datasets.}
We evaluate \textsc{UniRank} on two domain-specific multimodal retrieval tasks with hybrid text-image candidate sets.

\textbf{Scientific literature retrieval.}
We evaluate on the \textsc{MMDocIR} benchmark~\cite{dong2025mmdocir} and adopt its \emph{Academic Paper} domain under the \emph{layout-retrieval} setting.
Each document is a formal academic publication consisting of $m$ layout regions (e.g., paragraphs, figures, and tables) with structured layouts, citations, and academic visuals.
Given a natural-language question $Q$ (text-only), the system retrieves and reranks the top-$k$ most relevant layout regions from the paper.
This task naturally yields a heterogeneous candidate pool spanning multiple modalities.

\textbf{Design patent search.}
We additionally evaluate on a real-world design patent search task for product appearance risk screening.
Given a \emph{target product} as the query, the goal is to score each \emph{comparison patent} by whether it is \emph{substantially similar} in overall visual impression, and rerank the recalled patents accordingly.
Both queries and candidates are multimodal: a product includes a main image with optional title/description, while a patent contains structured fields such as patent title, representative figure, abstract, specification text, and metadata.
We construct the dataset by first retrieving candidate patents using an off-the-shelf appearance-based retrieval system, and then assigning binary labels (\texttt{high-risk}/\texttt{low-risk}) using \texttt{gemini-3-pro-preview}~\cite{team2023gemini}, which we treat as supervision for training and evaluation.

Dataset statistics for both domains, including the splits used for supervised fine-tuning ($\mathcal{D}_{\text{SFT}}$), hold-out preference construction ($\mathcal{D}_{\text{hold}}$), and the test set ($\mathcal{D}_{\text{test}}$), are summarized in Table~\ref{tab:dataset_stats}.
Additional qualitative examples and details of dataset construction and labeling are provided in Appendix~\ref{app:patent_data}.

\begin{table}[t]
\centering
\caption{Statistics of the two domain-specific datasets used in our experiments. All splits are disjoint.
% $\mathcal{D}_{\text{SFT}}$: SFT training data; $\mathcal{D}_{\text{hold}}$: held-out data for hard-negative mining; Test: final evaluation set.
}
\label{tab:dataset_stats}
\begin{tabular}{l l r r r}
\toprule
\textbf{Domain} & \textbf{Split} & \textbf{\# Docs} & \textbf{\# Queries} & \textbf{\# QA Pairs} \\
\midrule
\multirow{3}{*}{Scientific} 
    & $\mathcal{D}_{\text{SFT}}$    & 957 & 3980 & 71821 \\
    & $\mathcal{D}_{\text{hold}}$   & 239 & 995  & 17955 \\
    & $\mathcal{D}_{\text{test}}$   & 75  & 386  & 6840  \\
\midrule
\multirow{3}{*}{Patent} 
    & $\mathcal{D}_{\text{SFT}}$    & 8007 & 7891 & 79457 \\
    & $\mathcal{D}_{\text{hold}}$   & 998  & 986  & 9927  \\
    & $\mathcal{D}_{\text{test}}$   & 497  & 493  & 4953  \\
\bottomrule
\end{tabular}
\end{table}

\paragraph{Model and Training Framework.}
We instantiate \textsc{UniRank} using the \texttt{Qwen3-VL-8B-Instruct}~\cite{Qwen3-VL} as the backbone VLM, which supports interleaved text and image inputs and has demonstrated strong cross-modal reasoning capabilities.
Both the SFT and the subsequent GRPO-based RLHF stages are implemented within the \textsc{Swift} framework~\cite{zhao2025swift}, a modular toolkit for parameter-efficient fine-tuning of large models.
We employ LoRA~\cite{hu2022lora} to reduce memory overhead and accelerate training, while preserving the rich knowledge in the frozen VLM backbone.
Detailed hyperparameters and training configurations are reported in Appendix~\ref{app:hyperparams}.

\paragraph{Baselines.}
We compare \textsc{UniRank} against the following state-of-the-art reranking baselines:
\begin{itemize}
    \item \textbf{Text rerankers.}
    BGE-Reranker-v2-m3~\cite{chen2024bge} and Qwen3-Rera-nker~\cite{zhang2025qwen3}, two strong rerankers designed for text-only relevance scoring. 
    For multimodal inputs, we follow the common practice of converting non-text candidates (e.g., images/tables) into textual descriptions before reranking.

    \item \textbf{VLM list reranker.}
    MM-R5~\cite{xu2025mm}, a VLM-based listwise reranking method that homogenizes candidates by representing all items in an image-like format and directly generates a ranked list for the candidate set.

    \item \textbf{VLM pointwise rerankers.}
    jina-reranker-m0\footnote{\url{https://huggingface.co/jinaai/jina-reranker-m0}} and Qwen3-VL-Reranker~\cite{li2026qwen3}, two open-source VLM-based rerankers trained with large-scale data to output relevance scores for multimodal query--document pairs.
\end{itemize}

\paragraph{Evaluation metrics.}
We evaluate reranking quality using standard IR metrics, including Normalized Discounted Cumulative Gain (NDCG@k), Recall@k, and Mean Reciprocal Rank (MRR).
NDCG@k assesses ranked-list quality by assigning higher gain to relevant items at higher positions, with logarithmic discounting and normalization by the ideal ranking.
Recall@k indicates whether at least one ground-truth relevant candidate is retrieved within the top-$k$ positions.
MRR measures how early the first relevant item appears, computed as the average reciprocal rank of the highest-ranked relevant candidate over all queries.
Following common practice, we report results for $k \in \{1, 3, 5\}$.
Formal definitions of these metrics are provided in Appendix~\ref{app:metrics}.

\begin{table}[t]
\centering
\caption{Main results on scientific literature retrieval. For text rerankers, non-text layouts (images/tables) are converted into text using either OCR or VLM-generated descriptions provided by MMDocIR.}
\label{tab:main_science}
\resizebox{\linewidth}{!}{
\begin{tabular}{l|c|ccc|ccc|c}
\toprule
\multirow{2}{*}{Method} & \multirow{2}{*}{Size} &
\multicolumn{3}{c|}{Recall@k} &
\multicolumn{3}{c|}{NDCG@k} &
MRR \\
& & @1 & @3 & @5 & @1 & @3 & @5 &  \\
\midrule
BGE-Reranker-v2-m3 (OCR-text) & 0.6B & 26.7 & 40.0 & 57.8 & 26.7 & 30.5 & 38.1 & 38.3 \\
BGE-Reranker-v2-m3 (VLM-text) & 0.6B & 37.8 & 66.7 & 73.3 & 37.8 & 48.2 & 52.0 & 52.3 \\
Qwen3-Reranker (OCR-text)     & 8B   & 37.8 & 66.7 & 77.8 & 37.8 & 49.7 & 53.8 & 53.6 \\
Qwen3-Reranker (VLM-text)     & 8B   & 35.6 & 60.0 & 71.1 & 35.6 & 48.7 & 52.6 & 50.6 \\
MM-R5                         & 8B   & \underline{51.1} & 73.3 & 80.0 & \underline{51.1} & 59.1 & 63.0 & 63.4 \\
jina-reranker-m0              & 2B   & 48.9 & \underline{75.6} & \underline{88.9} & 48.9 & \underline{61.7} & \underline{67.1} & \underline{64.5} \\
Qwen3-VL-Reranker             & 8B   & 37.8 & 64.4 & 66.7 & 37.8 & 49.5 & 50.7 & 52.0 \\
\midrule
\textsc{UniRank} (ours)       & 8B   & \textbf{60.0} & \textbf{84.4} & \textbf{95.6} & \textbf{60.0} & \textbf{70.7} & \textbf{74.3} & \textbf{74.4} \\
\bottomrule
\end{tabular}}
\end{table}

\begin{table}[t]
\centering
\caption{Main results on design patent search. For text rerankers, we construct text-only inputs by concatenating all available textual fields.}
\label{tab:main_patent}
\resizebox{\linewidth}{!}{
\begin{tabular}{l|c|ccc|ccc|c}
\toprule
\multirow{2}{*}{Method} & \multirow{2}{*}{Size} &
\multicolumn{3}{c|}{Recall@k} &
\multicolumn{3}{c|}{NDCG@k} &
MRR \\
& & @1 & @3 & @5 & @1 & @3 & @5 &  \\
\midrule
BGE-Reranker-v2-m3      & 0.6B  & 27.6 & 59.2 & 76.3 & 27.6 & 36.9 & 44.6 & 48.5 \\
Qwen3-Reranker          & 8B    & 35.4 & 60.6 & 76.3 & 35.4 & 41.4 & 48.5 & 53.2 \\
MM-R5                   & 8B    & 52.1 & \underline{82.7} & 89.5 & 52.1 & 63.3 & 67.3 & 69.6 \\
jina-reranker-m0        & 2B    & 44.3 & 71.0 & 82.3 & 44.3 & 50.8 & 56.3 & 60.9 \\
Qwen3-VL-Reranker       & 8B    & \underline{57.9} & 81.6 & \underline{91.7} & \underline{57.9} & \underline{64.4} & \underline{70.0} & \underline{71.9} \\
\midrule
\textsc{UniRank} (ours) & 8B    & \textbf{65.2} & \textbf{83.5} & \textbf{93.4} & \textbf{65.2} & \textbf{70.8} & \textbf{74.9} & \textbf{76.5} \\
\bottomrule
\end{tabular}}
\end{table}

\subsection{Main Results}
\subsubsection{Effectiveness in domain-specific hybrid reranking (RQ1)}
\paragraph{\textbf{UniRank consistently outperforms strong text and VLM rerankers on hybrid candidates.}}
Table~\ref{tab:main_science} and Table~\ref{tab:main_patent} report the main reranking results on scientific literature retrieval and design patent search, respectively.
Across both domains, \textsc{UniRank} achieves the best overall performance on all evaluated metrics, demonstrating its effectiveness for domain-specific reranking over \emph{hybrid} text-image candidate sets.
Notably, \textsc{UniRank} improves both top-rank accuracy (Recall@1 / MRR) and ranking quality at larger cutoffs (NDCG@3/5), indicating better cross-modal score comparability and more robust ordering.

\textbf{Scientific literature retrieval.}
As shown in Table~\ref{tab:main_science}, text-only rerankers underperform in this hybrid setting even with image-to-text conversion.
Following MMDocIR, we report two conversion variants for non-text layouts:
\begin{enumerate*}[label=(\arabic*)]
    \item OCR-extracted text and 
    \item VLM-generated text descriptions.
\end{enumerate*}
Typically, VLM-generated descriptions often outperform OCR, but the gains are not consistent across rerankers, highlighting that modality conversion can be brittle due to variable description quality.
Among VLM-based baselines, MM-R5 is competitive but remains limited by its input homogenization strategy that treats candidates in an image-like format, which can be inefficient and may blur modality-specific signals.
Instead, \textsc{UniRank} natively scores mixed-modality candidates with a unified interface, and its domain-aligned training pipeline yields consistently better Recall@k, NDCG@k, and MRR, highlighting the benefit of end-to-end adaptation for scientific documents with complex layouts.

\textbf{Design patent search.}
Table~\ref{tab:main_patent} shows a similar trend in the design patent domain, where relevance is primarily determined by \emph{overall visual impression} and subtle appearance cues.
Text rerankers lag behind because cross-modal similarity cannot be reliably reduced to text, especially when fine-grained visual patterns dominate the judgment.
While open-source VLM rerankers provide stronger cross-modal reasoning, their general-domain training makes them less calibrated for domain-specific similarity criteria and metadata structure in patents.
By contrast, \textsc{UniRank} achieves the strongest performance, suggesting that combining hybrid-modality scoring with in-domain SFT and preference alignment is crucial for robust risk-oriented reranking.

\paragraph{\textbf{Efficiency advantage of native hybrid-modality scoring.}}
Beyond effectiveness, \textsc{UniRank} offers significant efficiency benefits by avoiding the need to convert text into images—a common practice in VLM-based rerankers like MM-R5. To quantify this, we measure
\begin{enumerate*}[label=(\arabic*)]
    \item \emph{storage overhead}: the total size of the candidate corpus when all items are represented as images, versus their native multimodal form; and 
    \item \emph{inference latency}: average time per query–candidate scoring on a single H20 GPU.
\end{enumerate*}
As shown in Table~\ref{tab:efficiency}, representing text as images inflates storage by 6.9$\times$ in the scientific domain and 1.4$\times$ in patents. Moreover, MM-R5’s image-heavy inputs lead to 2.6$\times$ and 2.8$\times$ slower inference than \textsc{UniRank}, which processes text natively and only encodes actual images. This confirms that \textsc{UniRank}’s hybrid-input design is not only more accurate but also far more practical for real-world deployment.

\begin{table}[t]
\centering
\caption{Efficiency comparison. Storage reflects the size of the candidate corpus when encoded in the method’s required input format. Inference latency measured on a single H20 GPU with batch size 1.}
\label{tab:efficiency}
\resizebox{\linewidth}{!}{
\begin{tabular}{l|cc|cc}
\toprule
\multirow{2}{*}{Method} &
\multicolumn{2}{c|}{Storage (GB)} &
\multicolumn{2}{c}{Inference Latency (ms)} \\
& Scientific & Patent & Scientific & Patent \\
\midrule
Qwen3-Reranker          & 0.7  & 0.2  &  7.6  & 9.8     \\
MM-R5                   & 83.5 & 6.0  & 42.0  & 1258.9  \\
\textsc{UniRank} (ours) & 12.1 & 4.2  & 16.2  & 446.4   \\
\bottomrule
\end{tabular}}
\end{table}

\begin{table}[t]
\centering
\caption{Robustness to first-stage retrievers on scientific literature retrieval. For each retriever, we report metrics without reranking (w/o) and with \textsc{UniRank} reranking (w/). ColQwen is the default retriever used in the main experiments.}
\label{tab:rq2_retrievers}
\resizebox{\linewidth}{!}{
\begin{tabular}{l|l|ccc|ccc|c}
\toprule
Retriever & Setting &
\multicolumn{3}{c|}{Recall@k} &
\multicolumn{3}{c|}{NDCG@k} &
MRR \\
&  & @1 & @3 & @5 & @1 & @3 & @5 &  \\
\midrule
BGE     & w/o & 24.4 & 40.0 & 40.0 & 24.4 & 30.4 & 30.9 & 33.2 \\
        & w/  & 48.9 & 75.6 & 77.8 & 48.9 & 59.0 & 61.2 & 62.5 \\
\midrule
DPR     & w/o & 15.6 & 31.1 & 35.6 & 15.6 & 21.7 & 23.0 & 24.2 \\
        & w/  & 46.7 & 71.1 & 73.3 & 46.7 & 57.6 & 59.2 & 59.7 \\
\midrule
ColBERT & w/o & 22.2 & 26.7 & 31.1 & 22.2 & 22.4 & 24.2 & 27.2 \\
        & w/  & 51.1 & 75.6 & 80.0 & 51.1 & 60.9 & 63.2 & 64.7 \\
\midrule
DSE     & w/o & 35.6 & 53.3 & 60.0 & 35.6 & 42.4 & 46.1 & 46.9 \\
        & w/  & 51.1 & 77.8 & 84.4 & 51.1 & 64.4 & 67.0 & 66.0 \\
% \midrule
% ColPali & w/o &  &  &  &  &  &  &  \\
%         & w/  &  &  &  &  &  &  &  \\
\midrule
ColQwen & w/o & 42.2 & 62.2 & 71.1 & 42.2 & 51.0 & 55.7 & 54.2 \\
        & w/  & 60.0 & 84.4 & 95.6 & 60.0 & 70.7 & 74.3 & 74.4 \\
\bottomrule
\end{tabular}}
\end{table}

\begin{table*}[t]
\centering
\caption{Robustness to backbone model size. We report results for \textsc{UniRank} instantiated with Qwen3-VL backbones of different sizes, alongside representative text and VLM reranker baselines at multiple scales.}
\small
\label{tab:rq2_scales}
\begin{tabular}{l|c|ccc|ccc|c|ccc|ccc|c}
\toprule
\multirow{2}{*}{Method} & \multirow{2}{*}{Size} &
\multicolumn{7}{c|}{Scientific} &
\multicolumn{7}{c}{Patent} \\
& &
\multicolumn{3}{c|}{Recall@k} &
\multicolumn{3}{c|}{NDCG@k} &
MRR &
\multicolumn{3}{c|}{Recall@k} &
\multicolumn{3}{c|}{NDCG@k} &
MRR \\
& & @1 & @3 & @5 & @1 & @3 & @5 &  & @1 & @3 & @5 & @1 & @3 & @5 &  \\
\midrule
Qwen3-Reranker          & 4B & 42.2 & 60.0 & 71.1 & 42.2 & 48.7 & 53.8 & 53.9 & 37.0 & 63.0 & 77.5 & 37.0 & 42.9 & 49.1 & 54.8 \\
                        & 8B & 37.8 & 66.7 & 77.8 & 37.8 & 49.7 & 53.8 & 53.6 & 35.4 & 60.6 & 76.3 & 35.4 & 41.4 & 48.5 & 53.2 \\
\midrule
Qwen3-VL-Reranker       & 2B & 33.3 & 60.0 & 62.2 & 33.3 & 44.3 & 45.9 & 46.7 & 43.9 & 72.6 & 85.5 & 43.9 & 52.3 & 58.9 & 61.5 \\
                        & 8B & 37.8 & 64.4 & 66.7 & 37.8 & 49.5 & 50.7 & 52.0 & 57.9 & 81.6 & 91.7 & 57.9 & 64.4 & 70.0 & 71.9 \\
\midrule
\textsc{UniRank} (ours) & 2B & 48.9 & 82.2 & \underline{93.3} & 48.9 & 63.6 & 68.1 & 66.8 & 62.2 & \textbf{85.3} & 91.8 & 62.2 & 68.6 & 73.4 & 75.0 \\
                        & 4B & \underline{53.3} & \underline{84.4} & 91.1 & \underline{53.3} & \underline{68.4} & \underline{70.2} & \underline{69.9} & \underline{65.0} & 83.5 & \underline{92.4} & \underline{65.0} & \underline{70.5} & \underline{74.3} & \underline{76.4} \\
                        & 8B & \textbf{60.0} & \textbf{84.4} & \textbf{95.6} & \textbf{60.0} & \textbf{70.7} & \textbf{74.3} & \textbf{74.4} & \textbf{65.2} & \underline{83.5} & \textbf{93.4} & \textbf{65.2} & \textbf{70.8} & \textbf{74.9} & \textbf{76.5} \\
\bottomrule
\end{tabular}
\end{table*}

\subsubsection{Robustness across retrieval setups and model scales (RQ2)}
\paragraph{\textbf{UniRank is robust across diverse first-stage retrievers.}}
We first study whether \textsc{UniRank} remains effective when the first-stage retriever changes, which alters the candidate distribution and difficulty.
On the scientific literature retrieval task, we consider five representative retrievers spanning both text and visual paradigms: three text retrievers (BGE~\cite{bge_embedding}, DPR~\cite{karpukhin2020dense}, and ColBERT~\cite{khattab2020colbert}) and two visual retrievers (DSE~\cite{ma2024unifying}, and ColQwen~\cite{faysse2024colpali}).
For text retrievers, we convert non-text layouts into OCR-extracted text.
As shown in Table~\ref{tab:rq2_retrievers}, adding \textsc{UniRank} on top of each retriever yields consistent gains across all metrics.
This suggests that \textsc{UniRank} does not overfit to a specific candidate distribution produced by a single retriever, but instead provides a stable and complementary relevance signal for hybrid candidates.
Notably, ColQwen yields the strongest initial retrieval performance, and thus is adopted as the default retriever in our main experiments. Even in this strong-retriever regime, \textsc{UniRank} still brings clear gains, indicating that its improvements are not merely due to fixing easy retrieval errors but also refining hard top-ranked cases.

\paragraph{\textbf{UniRank scales well with backbone capacity.}}
We further examine robustness to model capacity by instantiating \textsc{UniRank} with different Qwen3-VL backbone sizes.
We additionally report two representative baselines at multiple scales: the text-only Qwen3-Reranker (4B/8B) and the VLM-based Qwen3-VL-Reranker (2B/8B).
Table~\ref{tab:rq2_scales} summarizes results on both domains. For Qwen3-Reranker, we report its best result among the OCR-text and VLM-text conversions.
Overall, larger backbones consistently improve reranking performance, and \textsc{UniRank} remains the strongest method at each comparable scale.
In particular, \textsc{UniRank}-2B already matches or exceeds larger baseline rerankers in most metrics, while \textsc{UniRank}-8B achieves the best overall accuracy, indicating that our hybrid-modality scoring and end-to-end alignment recipe is effective across model sizes and provides a practical accuracy--cost trade-off.

\begin{table*}[t]
\centering
\caption{Ablation studies of \textsc{UniRank}. Each variant modifies one component of the full pipeline while keeping all other settings fixed. \textsc{UniRank}$^\ast$ denotes a variant that replaces our query-level grouping with the standard prompt-level grouping.}
\small
\label{tab:ablation}
\begin{tabular}{l|ccc|ccc|c|ccc|ccc|c}
\toprule
\multirow{2}{*}{Variant} &
\multicolumn{7}{c|}{Scientific} &
\multicolumn{7}{c}{Patent} \\
& \multicolumn{3}{c|}{Recall@k} & \multicolumn{3}{c|}{NDCG@k} & MRR
& \multicolumn{3}{c|}{Recall@k} & \multicolumn{3}{c|}{NDCG@k} & MRR \\
& @1 & @3 & @5 & @1 & @3 & @5 & 
& @1 & @3 & @5 & @1 & @3 & @5 & \\
\midrule
w/o SFT                     & 44.4 & 77.8 & 81.1 & 44.4 & 59.5 & 65.2 & 62.2 & 49.7 & 77.3 & 89.3 & 49.7 & 57.2 & 63.3 & 66.0 \\
w/o Hard-Negative Mining    & 51.1 & 80.7 & 80.9 & 51.1 & 63.6 & 65.3 & 66.8 & 63.6 & \underline{83.5} & \underline{93.4} & 63.6 & 69.5 & 73.9 & 75.5 \\
w/o RLHF                    & 46.7 & \underline{82.2} & 84.4 & 46.7 & 64.9 & 67.1 & 66.4 & 63.0 & \textbf{84.9} & 91.5 & 63.0 & 69.1 & 73.5 & 75.3 \\
\textsc{UniRank}$^\ast$     & \underline{53.3} & \underline{82.2} & \underline{88.9} & \underline{53.3} & \underline{65.9} & \underline{69.5} & \underline{67.9} & \underline{64.0} & 83.3 & \underline{93.4} & \underline{64.0} & \underline{70.0} & \underline{74.3} & \underline{75.8} \\
\midrule
\textsc{UniRank} (full)     & \textbf{60.0} & \textbf{84.4} & \textbf{95.6} & \textbf{60.0} & \textbf{70.7} & \textbf{74.3} & \textbf{74.4} & \textbf{65.2} & \underline{83.5} & \textbf{93.4} & \textbf{65.2} & \textbf{70.8} & \textbf{74.9} & \textbf{76.5} \\
\bottomrule
\end{tabular}
\end{table*}

\subsection{Ablation Studies (RQ3)}\label{sec:ablation}
This section ablates key components of \textsc{UniRank} to understand which design choices are essential for effective domain-specific reranking over hybrid text-image candidates. 
% We focus on the final reranking performance and report NDCG@k, Recall@k, and MRR on both domains.

\paragraph{Settings.}
Starting from the full \textsc{UniRank} pipeline, we consider the following controlled variants:
\begin{itemize}
    \item \textbf{w/o SFT:} skip supervised fine-tuning and directly perform preference-based training starting from the pretrained backbone.
    \item \textbf{w/o Hard-Negative Mining:} replace hard-negative mining with uniformly sampled negatives when constructing preference data, keeping the preference dataset size fixed.
    \item \textbf{w/o RLHF:} remove the entire preference-alignment stage and use the SFT model for reranking.
    \item \textbf{prompt-level GRPO (\textsc{UniRank}$^\ast$):} keep the full pipeline but replace our query-level grouping in GRPO with the standard prompt-level grouping (multiple responses per prompt), which ignores the listwise structure induced by sharing the same query.
\end{itemize}
All other settings (backbone, data splits, and inference protocol) follow the main experiments for fair comparison.

\paragraph{Results and analysis.}
Table~\ref{tab:ablation} summarizes the ablation results. Overall, the full \textsc{UniRank} performs best across both domains, and each component contributes in a complementary way.

\paragraph{\textbf{SFT is a necessary foundation for learning calibrated hybrid-modality scores.}}
Removing SFT leads to the largest performance drop on both datasets. 
This indicates that, for reranking framed as label-token scoring, instruction-following SFT is crucial to establish a usable scoring interface and to inject domain-specific relevance semantics before preference optimization.
Without this initialization, preference training becomes substantially less stable and less data-efficient, resulting in degraded ranking quality.

\paragraph{\textbf{Hard-negative mining substantially improves preference data quality.}}
Replacing hard negatives with random negatives consistently reduces performance.
Random negatives are often too easy and provide weak learning signals, whereas hard negatives capture the SFT model's in-domain failure modes (high-scoring but non-relevant candidates), producing more informative comparisons that improve the reward model and downstream GRPO optimization.

\paragraph{\textbf{Preference alignment further improves ranking robustness beyond SFT}}
Compared to the SFT-only variant (w/o RLHF), the full \textsc{UniRank} achieves consistent gains on Recall@1/MRR as well as NDCG@3/5, suggesting that RLHF improves both top-rank accuracy and overall ordering.
This supports that preference-based alignment helps correct systematic miscalibration of scores on difficult cross-modal cases, especially when candidates have heterogeneous modalities.

\paragraph{\textbf{Query-level grouping matters for reranking objectives.}}
\textsc{UniRank}$^\ast$ (prompt-level GRPO) underperforms our query-level GRPO despite using the same preference data.
This suggests that exploiting the listwise structure---contrasting candidates under the same query---provides a better training signal for reranking than treating each query--candidate prompt independently, leading to more consistent improvements in relative ordering.

\paragraph{\textbf{Summary.}}
The ablation results validate our end-to-end design: SFT provides a necessary domain-adapted and well-calibrated scoring interface; hard-negative mining supplies informative in-domain comparisons; RLHF further improves robustness and top-rank quality beyond SFT; and query-level GRPO better matches the listwise nature of reranking than prompt-level optimization.

\section{Conclusion}
In this paper, we presented \textsc{UniRank}, an end-to-end framework for training domain-specific rerankers over hybrid text-image candidate sets.
By leveraging the strong cross-modal alignment of VLMs, \textsc{UniRank} enables native and unified relevance scoring over mixed-modality inputs, preserving rich semantic cues while avoiding the overhead of converting text into images.
Our framework consists of two stages:
\begin{enumerate*}[label=(\arabic*)]
    \item instruction-driven SFT to learn calibrated relevance scoring, and 
    \item preference-based alignment with hard-negative mining and query-level GRPO to further refine ranking behavior using in-domain signals.
\end{enumerate*}
This pipeline allows \textsc{UniRank} to adapt effectively to specialized domains where both accuracy and efficiency are critical.

Extensive experiments show that \textsc{UniRank} consistently outperforms strong text-only and VLM-based reranking baselines, achieving higher Recall@k, NDCG@k, and MRR while maintaining lower storage and inference costs.
Ablation studies validate the contribution of each component, and robustness analyses demonstrate that \textsc{UniRank} generalizes well across diverse first-stage retrievers and backbone model sizes.

Future work includes extending \textsc{UniRank} to additional modalities (e.g., video and audio) and richer relevance signals (e.g., multi-grade labels), supporting multi-image candidates under larger context scales, and exploring fully unsupervised domain adaptation when labeled data is scarce.
We hope \textsc{UniRank} serves as a practical and scalable solution for domain-specific multimodal retrieval systems where hybrid candidate sets and resource constraints are prevalent.

\begin{acks}
This work was supported by Alibaba Group through Alibaba Innovative Research Program, the Science and Technology Commission of Shanghai Municipality (Grant No. 24510714300), and the Shanghai Municipal Science and Technology Major Project, China (Grant No. 2021SHZDZX0102).
\end{acks}

\bibliographystyle{ACM-Reference-Format}
\balance
\bibliography{sample-base}

\appendix

\newpage
\section{Evaluation Metrics}\label{app:metrics}
This section provides the formal definitions of the evaluation metrics used in our experiments, including Recall@k, NDCG@k, and Mean Reciprocal Rank (MRR).

\paragraph{Notation.}
For a query $q$, let the reranker output an ordered list of $n$ candidates $\pi_q = (c_1, \ldots, c_n)$, sorted by decreasing predicted relevance score.
Let $y_i \in \{0,1\}$ denote the ground-truth relevance label of candidate $c_i$ for query $q$ (1 for relevant, 0 for non-relevant).
Let $\mathrm{rank}_q$ denote the (1-indexed) position of the highest-ranked relevant candidate:
\begin{equation}
\mathrm{rank}_q = \min \{ i \mid y_i = 1 \},
\end{equation}
and $\mathrm{rank}_q = +\infty$ if no relevant candidate exists in the list.

\paragraph{Recall@k.}
Recall@k measures whether at least one relevant candidate is retrieved within the top-$k$ results:
\begin{equation}
\mathrm{Recall@}k(q) = \mathbb{I}\big[\min_{1\le i \le k} y_i = 1\big]
= \mathbb{I}\big[\mathrm{rank}_q \le k\big],
\end{equation}
where $\mathbb{I}[\cdot]$ is the indicator function.
We report the average Recall@k over all queries:
\begin{equation}
\mathrm{Recall@}k = \frac{1}{|\mathcal{Q}|}\sum_{q\in\mathcal{Q}} \mathrm{Recall@}k(q).
\end{equation}

\paragraph{MRR}
MRR measures how early the first relevant candidate appears:
\begin{equation}
\mathrm{RR}(q) =
\begin{cases}
\frac{1}{\mathrm{rank}_q}, & \mathrm{rank}_q < +\infty, \\
0, & \text{otherwise},
\end{cases}
\end{equation}
and Mean Reciprocal Rank averages this quantity over queries:
\begin{equation}
\mathrm{MRR} = \frac{1}{|\mathcal{Q}|}\sum_{q\in\mathcal{Q}} \mathrm{RR}(q).
\end{equation}

\paragraph{NDCG@k.}
NDCG@k evaluates ranked-list quality with position-based discounting.
We first define DCG@k for a query as:
\begin{equation}
\mathrm{DCG@}k(q) = \sum_{i=1}^{k} \frac{2^{y_i}-1}{\log_2(i+1)}.
\end{equation}
We then compute the ideal DCG@k by sorting candidates by ground-truth relevance:
\begin{equation}
\mathrm{IDCG@}k(q) = \max_{\pi} \sum_{i=1}^{k} \frac{2^{y^{\pi}_i}-1}{\log_2(i+1)},
\end{equation}
where $\pi$ ranges over permutations and $y^{\pi}_i$ is the relevance label at position $i$ under $\pi$.
Finally, NDCG@k is the normalized form:
\begin{equation}
\mathrm{NDCG@}k(q) =
\begin{cases}
\frac{\mathrm{DCG@}k(q)}{\mathrm{IDCG@}k(q)}, & \mathrm{IDCG@}k(q) > 0, \\
0, & \text{otherwise}.
\end{cases}
\end{equation}
We report mean NDCG@k across queries:
\begin{equation}
\mathrm{NDCG@}k = \frac{1}{|\mathcal{Q}|}\sum_{q\in\mathcal{Q}} \mathrm{NDCG@}k(q).
\end{equation}

\paragraph{Remarks.}
All metrics are computed on the reranked top-$k$ list produced by the reranker given the candidate set returned by the first-stage retriever.
In our experiments, we report $k\in\{1,3,5\}$.

\section{Dataset Construction Details}\label{app:patent_data}

\subsection{Scientific Literature Retrieval}\label{app:mmdocir}
We evaluate the scientific domain on the \textsc{MMDocIR} benchmark~\cite{dong2025mmdocir}, which targets multimodal information retrieval over long documents.
MMDocIR provides two retrieval granularities: page-level retrieval and fine-grained layout-level retrieval, where a \emph{layout} corresponds to a localized region such as a paragraph, figure, or table.
In this work, we focus on the \emph{layout-retrieval} setting in the \emph{Academic Paper} domain, where each document is a visually rich scientific paper with structured layouts, citations, and academic figures/tables.
Given a text-only question, the system retrieves and reranks the top-$k$ most relevant layout regions within the document.
As reported by MMDocIR, the evidence layouts in the Academic Paper domain span multiple modalities, including text, image, and table regions, making it a representative testbed for hybrid-candidate reranking.

\begin{figure*}[t]
    \centering
    % TODO: replace with your own example figure file
    \includegraphics[width=.85\linewidth]{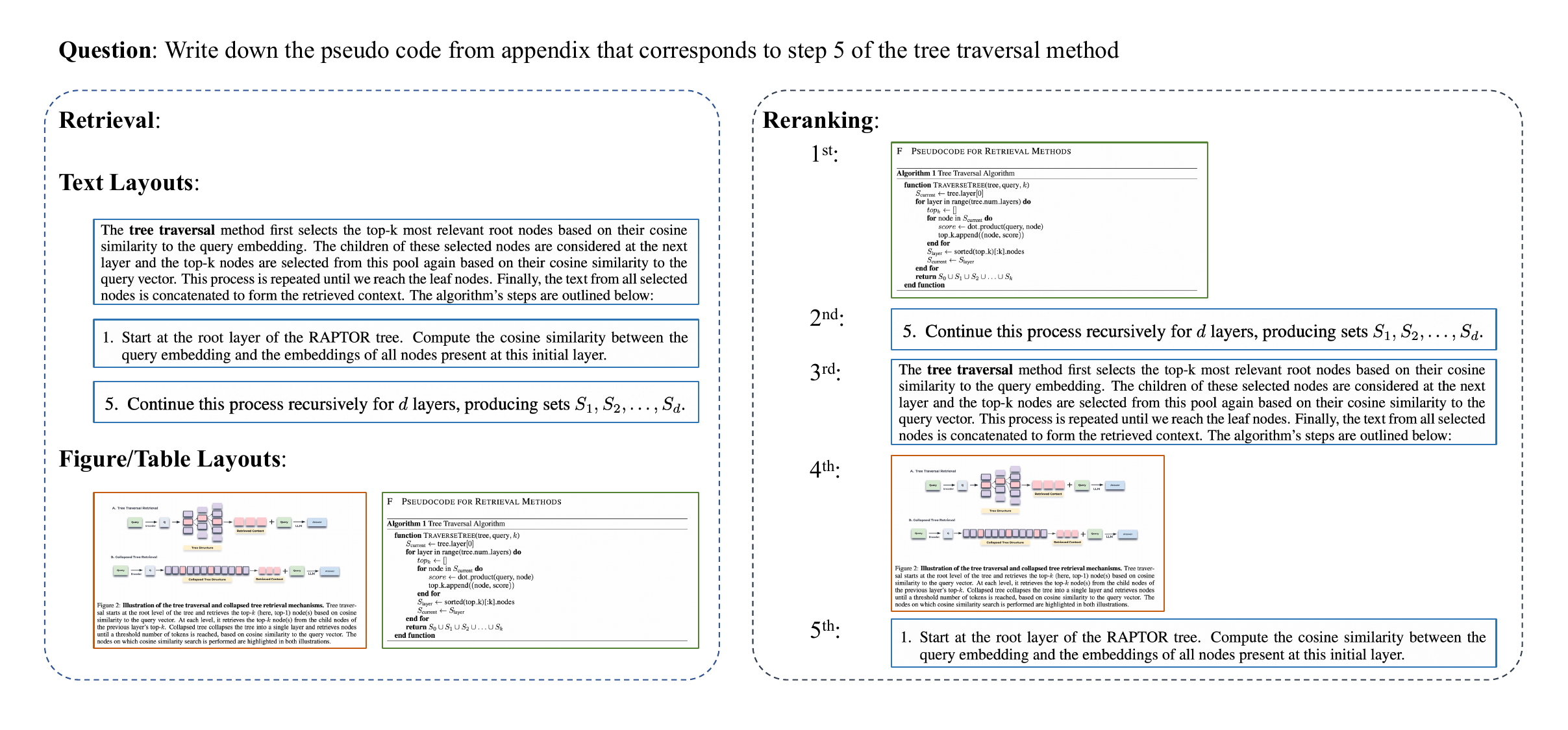}
    \caption{Example from MMDocIR Academic Paper (layout-retrieval). 
    Given a text-only question, the retriever/reranker selects the most relevant layout regions (e.g., paragraph/figure/table) within a long scientific document.}
    \label{fig:app_mmdocir_example}
\end{figure*}

For evaluation, we directly use the Academic Paper subset from the MMDocIR evaluation set.
Figure~\ref{fig:app_mmdocir_example} illustrates a representative layout-level reranking example, where a text-only query is matched against hybrid candidate regions (paragraphs, figures, and tables) within a scientific paper.

For training our reranker, we use the MMDocIR training set and focus on its \textsc{SciQAG} subset (scientific papers) to construct in-domain reranking supervision.
Specifically, we form query--candidate pairs from MMDocIR-provided layout-level labels and split them into the supervised fine-tuning set $\mathcal{D}_{\text{SFT}}$ and the hold-out set $\mathcal{D}_{\text{hold}}$ (used for hard-negative mining and preference data construction).
Additional examples of the instruction format and preference instances are provided in Appendix~\ref{app:prompt_examples}.

\begin{figure*}[t]
    \centering
    % TODO: replace with your own example figure file
    \includegraphics[width=.85\linewidth]{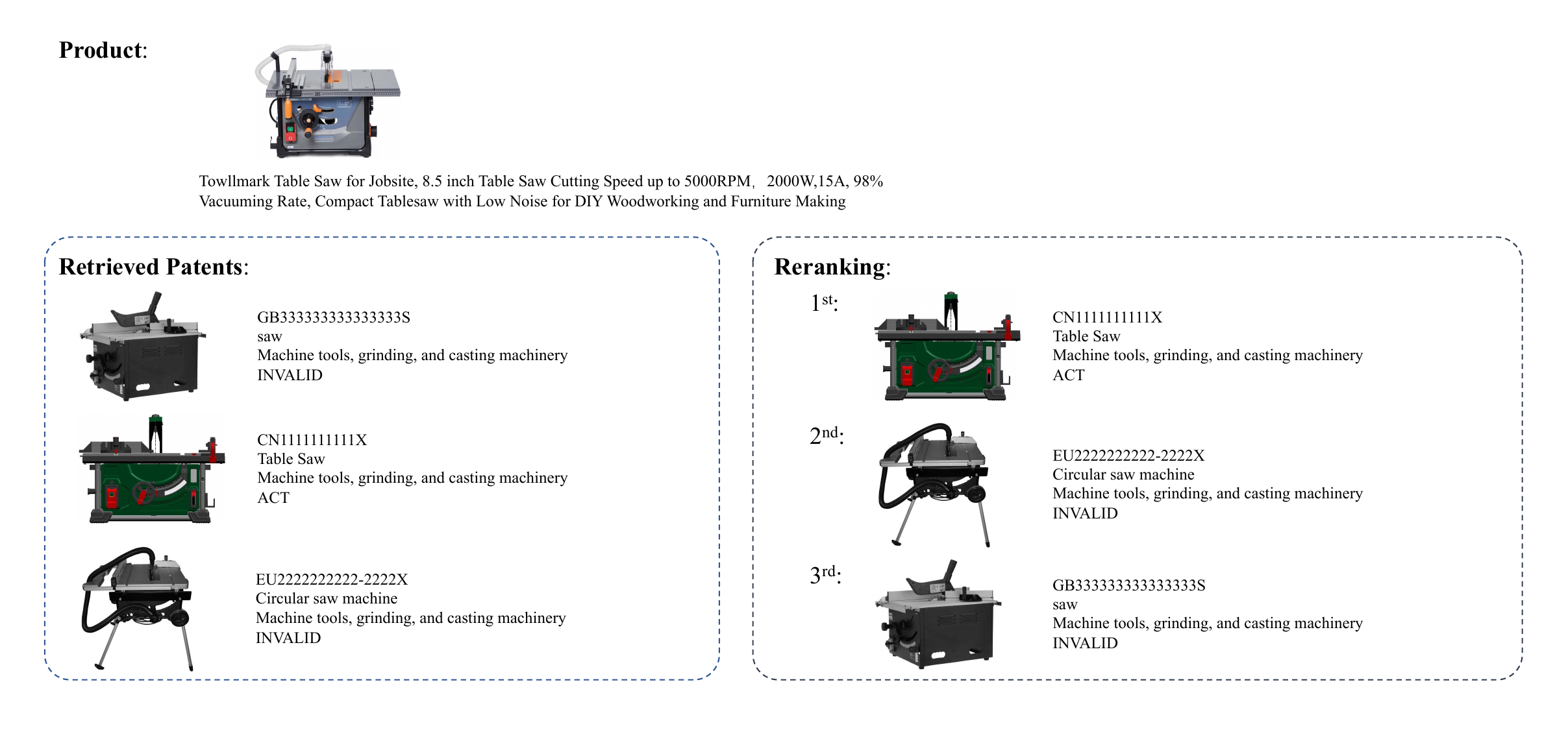}
    \caption{Example from the design patent search task. 
    Given a target product as the query (image + text), the reranker scores and orders multiple recalled design patents (candidates) by infringement risk.}
    \label{fig:app_patent_example}
\end{figure*}

\subsection{Design Patent Search}\label{app:patent}
\paragraph{Task definition.}
The design patent search task is motivated by real-world product appearance risk screening.
Given a \emph{target product} as the query and a set of recalled \emph{comparison patents} as candidates, the goal is to judge whether the product and each patent are \emph{substantially similar} in overall visual impression and output a reranking score for sorting the candidate patents.
Each query corresponds to multiple candidate patents, and the reranker is evaluated on its ability to place high-risk (similar) patents above low-risk (dissimilar) ones.

\paragraph{Candidate retrieval.}
For each target product, we first retrieve a candidate pool of potentially similar design patents using an off-the-shelf appearance-based retrieval system.\footnote{\url{https://open.eric-bot.com/docs}}
This step provides the initial candidate set for reranking and simulates the standard retrieve-then-rerank pipeline in deployment.

\paragraph{Labeling with a large multimodal model.}
To construct supervision signals for reranker training, we assign binary risk labels to each product--patent pair using \texttt{gemini-3-pro-preview}~\cite{team2023gemini}.
Concretely, for each query, we provide the model with
\begin{enumerate*}[label=(\arabic*)]
    \item the target product's main image and optional textual fields (title/description), and 
    \item each candidate patent's structured metadata (e.g., title, abstract, specification text, and patent identifier) together with representative patent figures.
\end{enumerate*}
The model is instructed to compare the two designs from an overall-appearance perspective and returns a structured prediction that includes a risk label.
We then treat the predicted risk label as the supervision signal for reranker training and evaluation.

\paragraph{Dataset construction for reranking.}
After labeling, we construct a standard reranking dataset where each query (product) is paired with its recalled candidates (patents) and corresponding risk labels.
We create the supervised fine-tuning set $\mathcal{D}_{\text{SFT}}$ from labeled query--candidate pairs, reserve a disjoint subset as $\mathcal{D}_{\text{hold}}$ for hard-negative mining and preference construction, and use a held-out test split for final evaluation.
Figure~\ref{fig:app_patent_example} illustrates the reranking setup in this domain, where a target product query is compared against multiple recalled design patents.
Examples of our system prompts, SFT instances, and preference pairs for this domain are provided in Appendix~\ref{app:prompt_examples}.

\newpage
\section{Implementation Details}\label{app:hyperparams}
This section summarizes the key implementation details of \textsc{UniRank}, including training hyperparameters and computational cost.

\subsection{Training Configuration}\label{app:train_config}
We implement both SFT and GRPO-based RLHF using the \textsc{Swift} framework~\cite{zhao2025swift}.
All experiments are conducted with parameter-efficient fine-tuning via LoRA~\cite{hu2022lora} on top of \texttt{Qwen3-VL-*} backbones.

\paragraph{Hyperparameters.}
Table~\ref{tab:app_hparams_sft} and Table~\ref{tab:app_hparams_grpo} list the hyperparameters used for SFT and GRPO, respectively.
Unless otherwise stated, we keep hyperparameters fixed across domains.

\begin{table}[t]
\centering
\caption{SFT hyperparameters.}
\label{tab:app_hparams_sft}
\resizebox{\linewidth}{!}{
\begin{tabular}{l l}
\toprule
\textbf{Hyperparameter} & \textbf{Value (Scientific / Patent)} \\
\midrule
Backbone model & Qwen3-VL-\{2B,4B,8B\}-Instruct \\
LoRA target modules & all-linear \\
LoRA rank $r$ & 16 \\
LoRA alpha & 32 \\
Learning rate & $5\times10^{-5}$ \\
LR schedule & linear \\
Warmup ratio & 0.05 \\
Epochs & 1 \\
Micro-batch size per GPU & 1 \\
Gradient accumulation steps & 8 \\
Global batch size & $8 \text{ GPUs} \times 1 \times 8 = 64$ \\
Max sequence length & 4096 / 16384 \\
Evaluation interval (steps) & 200 / 250 \\
Checkpoint save interval (steps) & 200 / 250 \\
% Dataloader workers & 8 \\
% Dataset preprocessing workers & 8 \\
\bottomrule
\end{tabular}}
\end{table}

\begin{table}[t]
\centering
\caption{RLHF hyperparameters.}
\label{tab:app_hparams_grpo}
\resizebox{\linewidth}{!}{
\begin{tabular}{l l}
\toprule
\textbf{Hyperparameter} & \textbf{Value (Scientific / Patent)} \\
\midrule
Policy init & SFT model \\
Reference policy & SFT model \\
LoRA target modules & all-linear \\
LoRA rank $r$ & 16 \\
LoRA alpha & 32 \\
Learning rate & $1\times10^{-6}$ \\
Warmup ratio & 0.05 \\
Epochs & 1 \\
Micro-batch size per GPU & 1 \\
Gradient accumulation steps & 1 \\
Global batch size & $8 \text{ GPUs} \times 1 \times 1 = 8$ \\
KL / ref regularization & 0.04 \\
Group size $G$ (query-level group\_n) & 30 / 5 \\
Sampling temperature & 1.0 \\
Top-$p$ & 0.9 \\
Max completion length & 4 / 16 \\
Truncation strategy & delete \\
vLLM for rollout & enabled, colocate mode \\
vLLM tensor parallel size & 4 \\
vLLM GPU memory utilization & 0.55 \\
Dataset shuffle & false \\
\bottomrule
\end{tabular}}
\end{table}

\paragraph{Hard-negative mining.}
For preference data construction, we mine hard negatives from the top-$N$ candidates ranked by the SFT reranker for each query. We set $N{=}30$ for scientific literature retrieval and $N{=}5$ for design patent search.

\paragraph{Inference settings.}
We use greedy decoding (temperature $=0$) and compute reranking scores from the label-token logits (Eq.~\ref{eq:score}).
For each query, we rerank the top-$k$ candidates returned by the first-stage retriever, where $k{=}100$ for scientific literature retrieval and $k{=}10$ for design patent search.

\subsection{Computational Cost}\label{app:compute_cost}
\paragraph{Hardware.}
All training runs are performed on a single compute node equipped with 8 NVIDIA H20 GPUs (96GB VRAM each), 192 CPU cores, and 1.8~TB RAM.

\paragraph{Approximate training time.}
Table~\ref{tab:app_training_time} reports the approximate wall-clock time for SFT, reward model (RM) training, and GRPO optimization.
Times may vary with backbone size, batch size, and the number of training steps.

\begin{table}[t]
\centering
\caption{Approximate training time of \textsc{UniRank} (in hours).}
\label{tab:app_training_time}
\resizebox{\linewidth}{!}{
\begin{tabular}{l c c c}
\toprule
\textbf{Domain} & \textbf{SFT} & \textbf{RM Training} & \textbf{GRPO (RLHF)} \\
\midrule
Scientific literature retrieval & 10.2 & 1.5 & 5.6 \\
Design patent search & 30.1 & 3.7 & 18.2 \\
\bottomrule
\end{tabular}}
\end{table}

\newpage
\section{Prompt Templates and Training Data Examples}\label{app:prompt_examples}
This section provides the prompt templates used in our experiments and illustrative examples of SFT instances and preference pairs for both domains. 

\subsection{Scientific Literature Retrieval}\label{app:prompt_science}
\paragraph{System prompt.}
We frame reranking as a binary relevance judgment over a question and a candidate layout region (text/table/figure). The system prompt is:
\begin{quote}
\small
\texttt{You are a multi-modal relevance judge.\\ Given a question and a document layout region (text/table/figure), determine whether this layout contains enough information to answer the question.\\ Respond only with 'yes' or 'no'.}
\end{quote}

\paragraph{SFT and preference data format.}
Each SFT instance is a single-turn chat example where the user message contains the query and one candidate (in its native modality, with images passed through the vision channel), and the assistant outputs exactly one label token (\texttt{yes} or \texttt{no}).
Preference data is constructed as pairwise comparisons under the same prompt context, where the chosen response is the ground-truth label and the rejected response is the opposite label.
Figure~\ref{fig:app_science_sft_pref} shows an illustrative SFT instance and the corresponding preference pair.

\begin{figure}[t]
    \centering
    \fbox{
    \parbox{0.95\linewidth}{
    \small
    \textbf{Scientific domain example.}\\[2pt]
    \textbf{(a) SFT instance}\\
    \textbf{$m^{\text{sys}}$:}
    See section \ref{app:prompt_science}.\\
    \textbf{$m^{\text{usr}}$:}
    \texttt{<QUERY>: Write down the pseudo code from appendix that corresponds to step 5 of the tree traversal method.}\\
    \texttt{<DOCUMENT>: step 5. Continue this process recursively for  $d$   layers, producing sets  $S_{1},S_{2},.,.,.,,S_{d}$.}\\
    \textbf{$m^{\text{asst}}$:}
    \texttt{yes}\\[6pt]
    \textbf{(b) Preference pair} \\
    \textbf{Prompt:} $m^{\text{sys}} + m^{\text{usr}}$\\
    \textbf{Chosen:} \texttt{yes} \quad \textbf{Rejected:} \texttt{no}
    }}
    \caption{Illustrative SFT and preference-data examples for scientific literature retrieval. The user message contains a question and one candidate layout region (text/table/figure). Images are provided via the VLM vision input channel.}
    \label{fig:app_science_sft_pref}
\end{figure}

\begin{figure}[t]
    \centering
    \fbox{
    \parbox{0.95\linewidth}{
    \small
    \textbf{Patent domain example.}\\[2pt]
    \textbf{(a) SFT instance}\\
    \textbf{$m^{\text{sys}}$:}
    See section \ref{app:prompt_patent}.\\
    \textbf{$m^{\text{usr}}$:}
    \texttt{<TARGET PRODUCT>: <image>; Title: Towllmark Table Saw for Jobsite; Description: 8.5 inch Table Saw Cutting Speed up to 5000RPM, 2000W, 15A ...}\\
    \texttt{<COMPARISON PATENT>: <image> Patent\_ID: CN1111111111X; Title: Circular saw machine; Abstract: Machine tools, grinding, and casting machinery ...}\\
    \textbf{Assistant:} \texttt{high-risk}\\[6pt]
    \textbf{(b) Preference pair}\\
    \textbf{Prompt:} $m^{\text{sys}} + m^{\text{usr}}$\\
    \textbf{Chosen:} \texttt{high-risk} \quad \textbf{Rejected:} \texttt{low-risk}
    }}
    \caption{Illustrative SFT and preference-data examples for design patent search. The user message contains one target product and one candidate patent with both images and structured text fields.}
    \label{fig:app_patent_sft_pref}
\end{figure}

\subsection{Design Patent Search}\label{app:prompt_patent}
\paragraph{System prompt.}
For the design patent domain, the reranker predicts an infringement-risk label under a common-observer criterion. The system prompt is:
\begin{quote}
\small
\texttt{You are a multimodal design-patent infringement risk reranker.\\ The input contains a TARGET PRODUCT (image + optional text + retrieval region) and a COMPARISON PATENT (patent figures + patent metadata).\\ Judge whether they are substantially similar in overall visual impression from a common-observer perspective.\\ Output exactly one label: high-risk or low-risk.\\ Do not output explanations or any extra characters.}
\end{quote}

\paragraph{SFT and preference data format.}
Each SFT instance compares one target product against one candidate patent. The user message includes the product image with optional title/description and the patent's representative figures plus structured metadata (e.g., patent identifier, title, abstract, and specification text).
The assistant outputs exactly one label: \texttt{high-risk} or \texttt{low-risk}.
Preference data is constructed analogously as pairwise comparisons under the same prompt context, with the ground-truth label as the chosen response.

Figure~\ref{fig:app_patent_sft_pref} provides an illustrative example of the SFT message structure and a preference pair.

\end{document}